\documentclass[aps,floatfix,superscriptaddress,reprint,10pt]{revtex4-1}
\usepackage[utf8]{inputenc}
\usepackage{mathrsfs}
\usepackage{amsmath}
\usepackage{amsfonts}
\usepackage{amssymb}
\usepackage{amsthm}
\usepackage{bm}
\usepackage{empheq}
\usepackage[normalem]{ulem}
\usepackage{multirow}
\usepackage{color}
\usepackage{graphicx}
\graphicspath{{./images/}}
\usepackage{geometry}
\usepackage{hyperref}
\usepackage{lineno}
\hypersetup{
     unicode=false,          % non-Latin characters in Acrobat’s bookmarks
     pdftoolbar=true,        % show Acrobat’s toolbar?
     pdfmenubar=true,        % show Acrobat’s menu?
     pdffitwindow=false,     % window fit to page when opened
     pdfstartview={FitH},    % fits the width of the page to the window
     pdftitle={My title},    % title
     pdfauthor={Author},     % author
     pdfsubject={Subject},   % subject of the document
     pdfcreator={Creator},   % creator of the document
     pdfproducer={Producer}, % producer of the document
     pdfkeywords={keyword1} {key2} {key3}, % list of keywords
     pdfnewwindow=true,      % links in new PDF window
     colorlinks=false,       % false: boxed links; true: colored links
     linkcolor=red,
     citecolor=green,        % color of links to bibliography
     filecolor=magenta,      % color of file links
     urlcolor=cyan           % color of external links
}
\usepackage[toc,page]{appendix}

\begin{document}

% Title
\title{Interplay of Wave Localization and Turbulence in Spin Seebeck Effect}
\author{C. Wang}
\affiliation{School of Electronic Science and Engineering
and State Key Laboratory of Electronic Thin Film and
Integrated Devices, University of Electronic Science and Technology
of China, Chengdu 610054, China}
\author{Yunshan Cao}
\affiliation{School of Electronic Science and Engineering
and State Key Laboratory of Electronic Thin Film and
Integrated Devices, University of Electronic Science and Technology
of China, Chengdu 610054, China}
\author{X. R. Wang}
\email[Corresponding author: ]{phxwan@ust.hk}
\affiliation{Physics Department, The Hong Kong University of Science and Technology,
 Clear Water Bay, Kowloon, Hong Kong}
\affiliation{HKUST Shenzhen Research Institute, Shenzhen 518057, China}
\author{Peng Yan}
\email[Corresponding author: ]{yan@uestc.edu.cn}
\affiliation{School of Electronic Science and Engineering
and State Key Laboratory of Electronic Thin Film and
Integrated Devices, University of Electronic Science and Technology
of China, Chengdu 610054, China}
\date{\today}

\begin{abstract}
One of the most important discoveries in spintronics is the spin
Seebeck effect (SSE) recently observed in both insulating and (semi-)conducting magnets.
However, the very existence of the effect in transverse configuration is still a subject of current debates, due to conflicting results reported in different experiments. Present understanding of the SSE is mainly based on a
particle-like picture with the local equilibrium approximation (LEA), i.e., spatially resolved temperature-field assumed to describe the system. In this work, we abandon the LEA to some extent and develop a wave theory to explain the SSE, by highlighting the interplay between wave localization and turbulence.
We show that the emerging SSE with a sign change in the high/low-temperature regions is closely related
to the extendedness of the spin wave that senses an average temperature
of the system. On the one hand, ubiquitous disorders (or magnetic field gradients) can strongly suppress the transverse spin Seebeck effect (TSSE) due to Anderson (or Wannier-Zeeman) spin-wave localization. On the other hand, the competing wave turbulence of interacting magnons tends to delocalize
the wave, and thus remarkably revives the TSSE before the magnon self-trapping. Our theory provides a promising route to resolve the heated debate on TSSE with a clear experiment scheme to test it in future spin caloritronic devices.
\end{abstract}

\maketitle

\section{Introduction}

Spin Seebeck effect (SSE) refers to the generation of
a spin voltage by temperature gradient in magnetic materials \cite{Uchida}.
There are two typical experimental configurations to measure the SSE:
One is the transverse SSE (TSSE)
\cite{Uchida,Jaworski2,Uchida1,McLaughlin1,Huang1,Schmid1,Meier1,Bui1,Soldatov1,Meier}
in which a nonmagnetic
metal bar is vertically placed on top of the magnetic strip.
The other is the longitudinal SSE (LSSE) \cite{LSSE1,LSSE2,LSSE3}
where the corresponding metal bar is connected to one end of the
magnetic strip, and the thermal gradient is perpendicular to the
magnetic$|$nonmagnetic interface. While the
LSSE is less questionable, the existence of the TSSE is
a subject of recent debates (see Table~\ref{table1} for summary)
\par

The present interpretation of SSE is mainly based on a ``particle-like" theory \cite{Xiao,Adachi2,Hiroto,Jaworski1}
pioneered by Sanders and Walton \cite{Sanders}, assuming that both the magnon-magnon and phonon-phonon
interactions are much stronger than the magnon-phonon
coupling so that magnons and phonons form two coupled
Boson gasses with different local temperatures \cite{Xiao,Adachi2}.
The local energy exchange rate between the two gasses is proportional to their
temperature difference. However, the measured temperature difference
between magnon and phonon is too small \cite{Agrawal}
to explain the TSSE. An alternative view relies on the phonon-drag mechanism
\cite{Hiroto,Jaworski1} that aims to explain the enhancement of the TSSE
signal at low temperature. This mechanism should be important for a strong
magnon-phonon interaction, which is not consistent with experimental evidences of relatively weak magnon-phonon coupling in large-scale
transports in magnetic insulator \cite{Agrawal,Guo1}.
%{\color{blue} In summary, the particle-like approach \cite{Xiao,Adachi2,Hiroto,Jaworski1}
%can explain the observed sign reversal of the spin current
%at higher and lower temperature regions in some TSSE experiments \cite{Uchida1},
%but fail to explain why the similar phenomena can not be observed in different
%experiments \cite{Meier}, as well as certain key experimental
%aspects \cite{Agrawal,Guo1}.}
\par

\begin{table*}[!ht]\centering
\caption{\label{table1}The table summarizes experimental results of
the TSSE in ferro-(or ferri-)magnetic metals (FMs),
semiconductors (FSCs), and insulators (FIs), as well as the length of
the sample used in corresponding measurements. In these experiments, the
length of the thin films $L$ ranges from 5~mm to 8~mm. Conventionally, the TSSE is detected through the
inverse spin Hall effect in the adjacent heavy metal bar. Thus,
in early experiments, the electric detections are inevitably
mixed with the thermoelectric and magnetoelectric effects.
A recent experiment \cite{McLaughlin1} reports the TSSE via
a novel optical method.
}
\begin{ruledtabular}
\begin{tabular}{cccc}
Reports of TSSE & Length $L$ (mm) &
Reports of no TSSE & Length $L$ (mm) \\
\hline
Uchida, K. {\it et al} \cite{Uchida} (FM) &    8 &
Huang, S. Y. {\it et al} \cite{Huang1} (FM) & 5 \\
Jaworski, C. M. {\it et al} \cite{Jaworski2} (FSC) & 8 &
Schmid, M. {\it et al} \cite{Schmid1} (FM) & 5  \\
Uchida, K. {\it et al} \cite{Uchida1} (FI) & 5 &
Meier, D. {\it et al} \cite{Meier1} (FM) & 5  \\
McLaughlin R. {\it et al}  \cite{McLaughlin1} (FM) & 5 &
Bui, C. T. {\it et al} \cite{Bui1} (FM) & 5  \\
 & &
Soldatov, I. V. {\it et al} \cite{Soldatov1} (FSC) & 5  \\
 & &
Meier, D. {\it et al} \cite{Meier} (FI) & 5  \\
\end{tabular}
\end{ruledtabular}
\end{table*}

Two assumptions are often adopted in the particle-like theory:
First, spin waves in ferromagnets are assumed to be extended although a typical
``propagation length'' of thermal magnons is also introduced
phenomenologically in clean systems.
However, inevitable material imperfections introduce 
disorders \cite{Sun1,Wesenberg1}, which
may lead to the spin-wave Anderson localization \cite{Nowak}.
Meanwhile, the applied temperature gradient can generate a spatial
dependence of the magnetic anisotropy \cite{Antropov,Garad,Hannes1},
so that local magnetic moments experience an inhomogeneous magnetic field
which can result in a spin-wave Wannier-Zeeman localization \cite{Yan1}.
Second, magnons are assumed to be in thermal equilibrium with the
local surroundings such that a position-dependent magnon temperature is
introduced, while the global system is still out of equilibrium. Nonetheless,
this local equilibrium approximation (LEA) often leads to contradictions,
e.g., the local temperature obtained by the kinetic method
is different from that by the entropic approach \cite{Narayanan1}.
Furthermore, the SSE should not require phonons for
existence, just like its counterpart of normal Seebeck effect where
phonon usually plays a bad role, and one tries to eliminate its effect.
Thus, it is highly interesting and important to see whether one
can understand the spin transport in the TSSE with neither the LEA nor the magnon-phonon coupling (at least not explicitly),
which motivates us to revisit this outstanding issue and to propose a more
general wave theory.
\par

\begin{figure}[ht!]
\centering
  \includegraphics[width=0.48\textwidth]{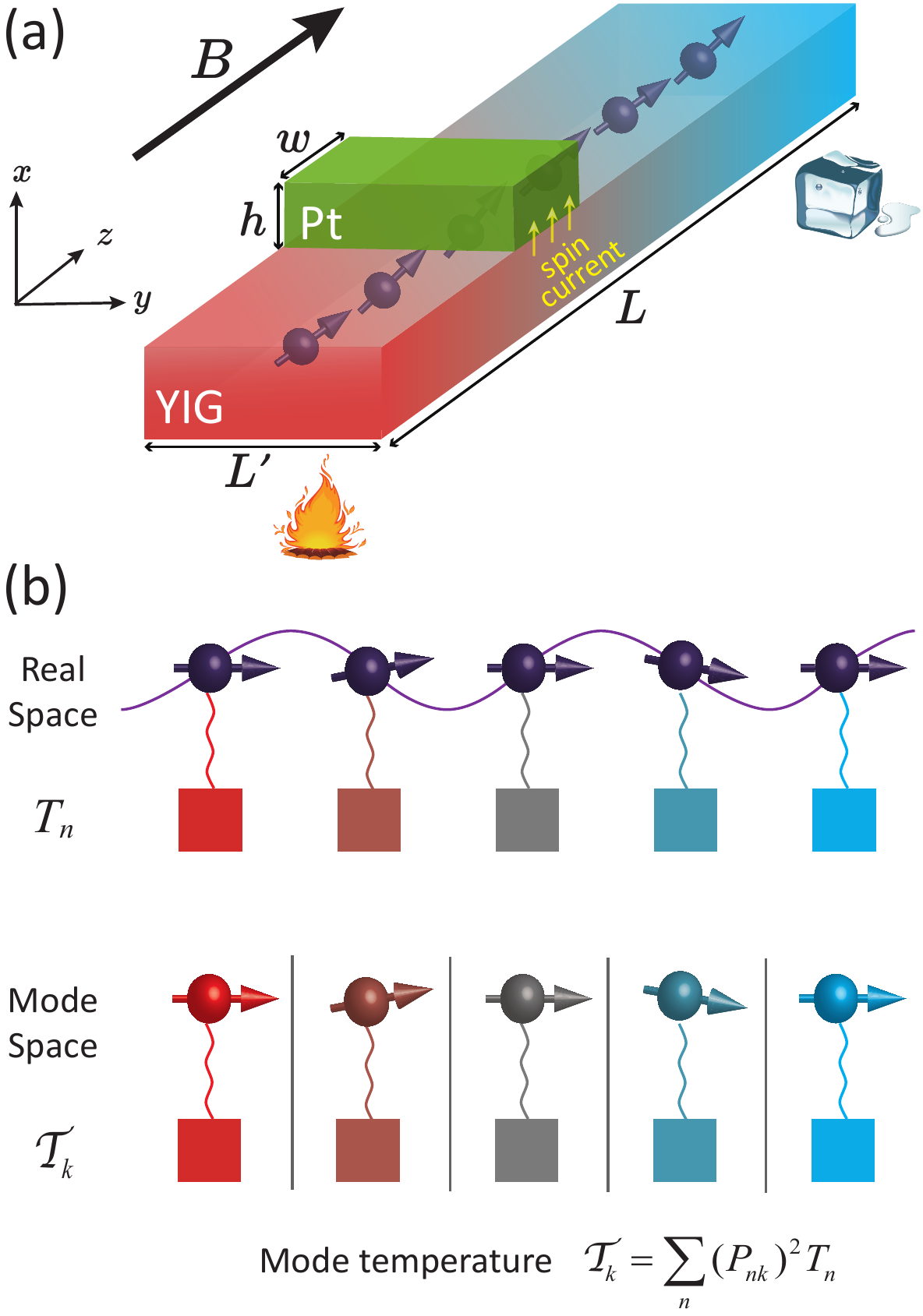}
  \caption{(a) Schematic plot of the YIG$|$Pt system, which
  includes a rectangle-shaped YIG with a Pt detect bar attached
  on its surface. A temperature gradient $\nabla T$ is
  applied in the quasi 1D system along $z-$direction. The net spin current
  is injected along the perpendicular direction with spatial distributions $j^z_{s,n}$,
  which is converted into an electric voltage by the inverse spin Hall effect.
  For arbitrary temperature profile
  $T_n$, $j^z_{s,n} \propto  \left(\sum_k (P_{nk})^2 \mathcal{T}_k-T_n\right)$,
  with $\mathcal{T}_k$ the mode temperature of magnons and $P_{nk}$ the
  spin wave function.
  (b) Temperature in both the real
  and the mode spaces. According to the principle of energy repartition at nonequilibrium steady states
  \cite{Yan1}, magnon at a given mode $k$ carries
  an energy $E_k=k_B \mathcal{T}_k$ with $\mathcal{T}_k=\sum_n (P_{nk})^2 T_n$.
  Thus, the nonequilibrium system can be decomposed into
  ``equilibrium'' subsystems with well-defined spectrum temperature $\mathcal{T}_k$
  in the normal-mode space.
  }
\label{fig_schmatic}
\end{figure}

\begin{figure*}
\centering
  \includegraphics[width=0.88\textwidth]{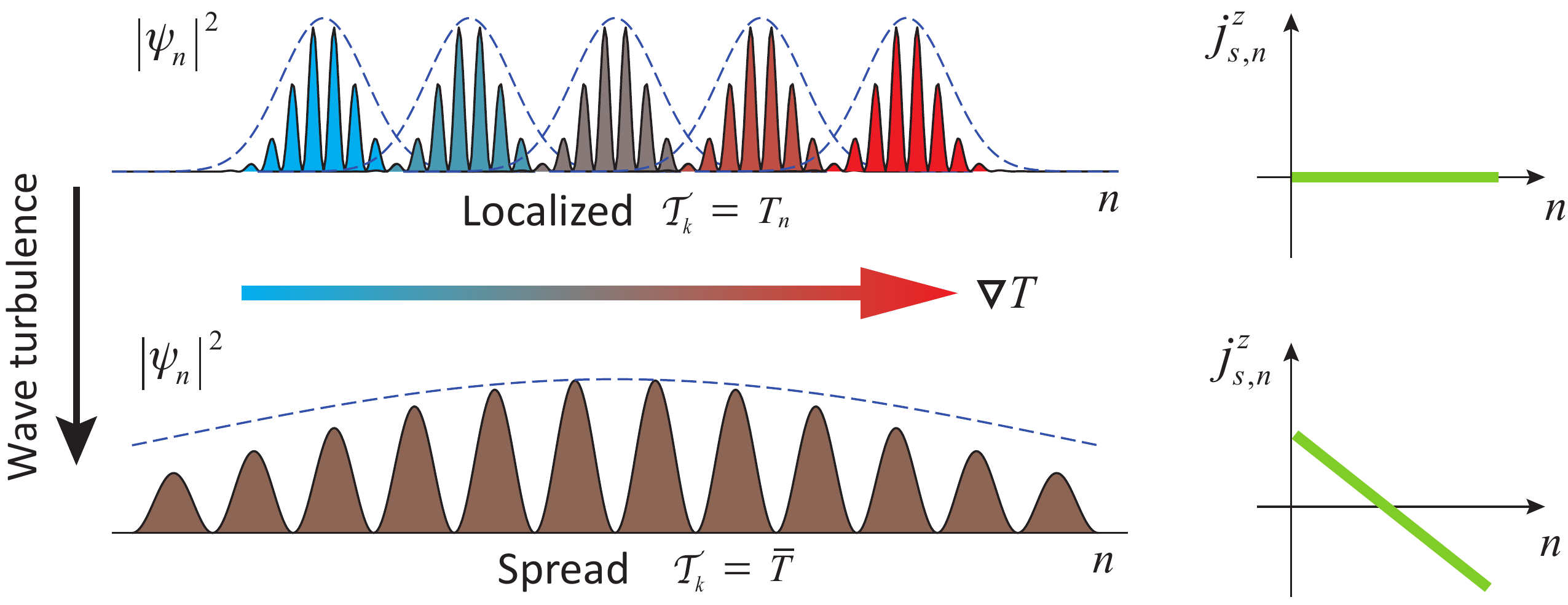}
  \caption{The TSSE incorporates three physical situations:
  (i) For highly localized spin waves, $P_{nk}\simeq\delta_{nk}$, local magnons
  and electrons reach thermal equilibrium. In this case, $\mathcal{T}_k=T_n$,
  and no TSSE. (ii) For spread spin waves, magnons act globally
  such that the temperature for all modes is equal to the average temperature of the
  YIG, $\mathcal{T}_k =\bar{T}$. In this case, $j^{z}_{s,n}\propto\bar{T}-T_n$,
  and one should be able to observe the TSSE.
  (iii) Although the presence of disorders and field-gradients in disordered
  and/or anisotropic YIG tends to localize the spin wave, the intrinsic nonlinear
  nature of spin waves competes with the localizations. At moderate magnon-magnon
  interactions, spin waves can be delocalized, and thus revive the TSSE.
  }
\label{fig_schmatic1}
\end{figure*}

In this article, we use the stochastic Landau-Lifshitz-Gilbert
(LLG) equation to describe the magnetization dynamics and to formulate the interplay between spin-wave localization and
turbulence \cite{Nazarenko} which captures the statistical property of large numbers of incoherent interacting magnons. The effects of phonons and
other possible degrees of freedom such as electromagnetic
radiations are to provide a stochastic magnetic field \cite{Andreas,Photon} and
to establish a local temperature for electrons in the
heavy-metal detecting bar. We consider the Y$_3$Fe$_5$O$_{12}|$Pt (YIG$|$Pt)
bilayer as the model system [see Fig.~\ref{fig_schmatic}(a)]. We show that the TSSE is a consequence of the extendedness of spin
waves and the temperature of nonequilibrium magnons is mode dependent. Three relevant length scales are identified: the localization length of spin
waves $\xi$, the length of YIG film $L$,
and the width of Pt contact $w$. Spread spin waves ($\xi\sim L$) can sense an average
temperature of the system, and are therefore not
in thermal equilibrium with electrons in the Pt contact. The TSSE is thus generated with a sign change in the high-/low-temperature regions, detected by the inverse spin Hall voltage. On the contrary, highly-localized spin waves ($\xi\sim w$) sense
only the local bath temperature and are in thermal equilibrium with the itinerant
electrons, leading to a strong suppression of the TSSE.
Our idea is schematically shown in Fig.~\ref{fig_schmatic1}.
Interestingly, we find that magnon-magnon interactions play remarkable roles in the spin transport: (i) At elevated temperature, spin-wave
turbulence competes with the localization to make magnons more
extended, and therefore revive the TSSE, and (ii) Magnon with a renormalized dispersion relation serves as the spin and heat carrier.
\par

\section{Wave theory of TSSE}

We model the YIG film by a two-dimensional (2D) Heisenberg ferromagnet 
of length $L$ and width $L'$ with $N=LL'/a^2$ local
magnetic moments $\vec{S}_{\bm{n}}=S\vec{s}_{\bm{n}}$ where $S$ and
$\vec{s}_{\bm{n}}$ are respectively the magnitude and the unit vector of the magnetic moment at
site $\bm{n}=(n_1,n_2)$ (both $n_1$ and $n_2$ are integers, and $1\leq
n_1\leq L'/a$, $1\leq n_2\leq L/a$). Here $a=1.24~\hbox{nm}$ is the lattice
constant. We start from the following Hamiltonian
\begin{equation}
    \begin{gathered}
        \mathcal {H}=-\sum_{\langle\bm{n},\bm{n'}\rangle}J\vec{s}_{\bm{n}}
        \cdot\vec{s}_{\bm{n'}}+\sum_{\bm{n}}[D_{\bm{n}}(s^z_{\bm{n}})^2-BSs^z_{\bm{n}}],
    \end{gathered}
\end{equation}
where $J>0$ is the nearest-neighbour exchange coupling constant and $D_{\bm{n}}$
is the local anisotropy constant. An external magnetic field $B\textbf{e}_z$ is applied to align the spin
along $z-$direction [see Fig.~\ref{fig_schmatic} (a)]. Each spin is
connected with a local Langevin thermal bath with temperature $T_{\bm{n}}$.
To generate the spin current
pumped from the YIG to Pt contact, we consider a linear temperature profile
$T_{\bm{n}}=T_0+\Delta T(n_2 a/L)$ in YIG with $T_0=300~\hbox{K}$ and
$\Delta T=20~\hbox{K}$, without loss of generality.
\par

The spin dynamics is then governed by the
stochastic LLG equation \cite{Hoffman}. For small spin fluctuations
$\vec{s}_{\bm{n}}=(s^x_{\bm{n}},s^y_{\bm{n}},1)$ with
$|s^{x,y}_{\bm{n}}|\ll 1$, the stochastic
LLG equation reduces to the following stochastic dissipative discrete nonlinear
Schr\"{o}dinger equation for
$\psi_{\bm{n}}\equiv s^x_{\bm{n}}+is^y_{\bm{n}}$
\begin{equation}
    \begin{gathered}
        (i+\alpha)\dot{\psi}_{\bm{n}}=\dfrac{\gamma}{S}\left[J
        (\psi_{n_{1},n_{2}+1}
        +\psi_{n_{1},n_{2}-1})\right.\\
        +J(\psi_{n_{1}+1,n_{2}}
        +\psi_{n_{1}-1,n_{2}})
        \left.-K_{\bm{n}}\psi_{\bm{n}}-\nu_{\bm{n}}|\psi_{\bm{n}}|^2\psi_{\bm{n}}\right]+\Theta_{\bm{n}},
    \end{gathered}\label{DNLS}
\end{equation}
where $\alpha$ is the dimensionless Gilbert damping constant, $\gamma$ is the
gyromagnetic ratio, and $K_{\bm{n}}=4J+BS-2D_{\bm{n}}$.
An anisotropy-induced nonlinear four-magnon term,
i.e., $-\nu_{\bm{n}}|\psi_{\bm{n}}|^2\psi_{\bm{n}}$ with $\nu_{\bm{n}}=D_{\bm{n}}$,
is considered \cite{Kosevich}. To model disorders and field gradients, we assume
that the ``on-site energy'' $K_{\bm{n}}$ is random and uniformly distributed
in the range of $[H_0+\epsilon n_2-W/2,H_0+\epsilon n_2+W/2]$ on top of a
constant field $H_0$, with $\nu_{\bm{n}}=\nu=D$.
Here $W$ and $\epsilon$ measure the strength of the disorder and the field gradient,
respectively.
In the following discussions, both $K_{\bm{n}}$ and $\nu$ are measured in
the unit of $J$. Finite temperature effect is modeled by stochastic fields
$\Theta_{\bm{n}}$ satisfying the fluctuation-dissipation theorem \cite{Brown1},
$\langle \Theta_{\bm{n}} (t)\Theta^{\ast}_{\bm{n}'}(t') \rangle=
(4\alpha k_B T_{\bm{n}}\gamma/S) \delta_{n_1 n'_1}\delta_{n_2 n'_2}\delta(t-t')$, where
$k_B$ is the Boltzmann constant. We set $k_B=1$ such that
the energy has the same unit with the temperature.
\par

Disorders and inhomogeneous magnetic fields lead to the Anderson
and Wannier-Zeeman localizations of spin waves, respectively.
For $\nu=\epsilon=0$ and $W\neq 0$ (disorder only),
the effective spin-wave Hamiltonian described by Eq.~\eqref{DNLS}
belongs to the Gaussian orthogonal ensemble, and all spin waves are
exponentially localized (Anderson localizations) \cite{Wang123}. While for
$\nu=W=0$ and $\epsilon\neq 0$ (field gradient
only), magnon states are periodic along $y-$direction and 
Wannier-Zeeman localized along $z-$direction with a localization length $\xi\simeq -a/[(\epsilon/J)\ln(\epsilon/J)]$ for $\epsilon\rightarrow 0$ \cite{Yan1,Kolovsky}.
\par

In the presence of a temperature gradient, the spin current is pumped into the
Pt contact from the locally fluctuating spins in YIG, and its
DC component at site $\bm{n}$ is given by \cite{Jiao1,Chen1}
\begin{equation}
        \begin{gathered}
        j^z_{s,\bm{n}}=-\dfrac{g^{\uparrow\downarrow}_{\text{eff}}\hbar}{4\pi}
        \Bigg\langle\text{Im}[\psi_{\bm{n}}\dot{\psi}_{\bm{n}}] \Bigg\rangle,
        \end{gathered}\label{current1}
\end{equation}
where $g^{\uparrow\downarrow}_{\text{eff}}$ is the effective spin mixing
conductance at the interface, and $\langle\cdots\rangle$ denotes time average.
Note that an accompanied spin current from the Pt contact back
into YIG, which is proportional to the electron temperature $T_e$ at the Pt 
contact \cite{Foros}, must be taken into account to obtain the net spin current.
Disregarding the Kapitza interfacial heat resistance, the electrons at
the Pt contact are in thermal equilibrium with the local heat bath, i.e., $T_e=T_{\bm{n}}$.
\par

We first focus on the limit of (quasi) 1D spin chain ($n_{1}=1,n_{2}=n$) since analytical solutions of the problem are likely to obtain. We are then able to get the expression of the net interfacial spin current
\begin{equation}
\begin{gathered}
j^z_{s,n}=\dfrac{g^{\uparrow\downarrow}_{\text{eff}}\hbar\gamma}{2\pi S}
\left(\sum_k (P_{n k})^2\mathcal{T}_k-T_{n}\right),
\end{gathered}\label{current}
\end{equation}
where $P_{n k}$ is the amplitude of wave function of the $k$th magnon
mode at site $n$, and $\mathcal{T}_k\equiv\sum_{n}(P_{n k})^2T_{n}$ is the corresponding
mode temperature under the bath temperature field $T_n=T_0+\Delta T(na/L)$. Detailed derivations are presented in the Appendix~\ref{app1}. It reveals one of the main
differences between the present wave theory and the popular
particle-like one \cite{Xiao,Adachi2,Hiroto,Jaworski1,Sanders}:
Instead of introducing a position-dependent magnon temperature
$T^{\text{Magnon}}_{n}$ using the LEA, we show that the temperature of non-equilibrium
magnons is genuinely spectrum resolved as $\mathcal{T}_k$.
The origin of TSSE then can be understood as follows: For extended spin waves
with $\xi\simeq\infty$, they sense an average temperature of the whole system
$\mathcal{T}_k=\bar{T}=\sum_{n} T_na/L$, such that the magnon temperature in the
high- (low-)$T$ region is lower (higher) than the bath temperature.
The sign of TSSE is, therefore, opposite at the two ends.
We emphasize that extended spin waves generally do not exist in disordered
1D spin chain since all spin waves are localized with a localization length
$\xi\simeq (96J^2/W^2)a$ \cite{kramer1}. However, as we illustrate later, as far as spin waves
with long enough localized length ($\xi\sim L$ for spread spin waves), the magnons
can still sense a mean temperature of the YIG.
In the opposite limit, however, highly-localized magnons ($P_{nk}\sim\delta_{nk}$) inherit the same local
temperature as the itinerant electrons in the Pt contact, and therefore cannot generate a net
spin current, consistent with the second law of thermodynamics \cite{yuhua}. Our arguments are qualitatively applicable to 2D, although the corresponding localization length is longer by a factor from 1 to 3  than 1D (see below).
\par

\section{Anderson and Wannier-Zeeman localizations}

To verify our analytical result, we numerically calculate the inverse
spin Hall voltage $V_H(z)$ detected in the Pt contact of the YIG$|$Pt
systems (See Methods~\ref{method-1} and Supplemental Material Sec.~I \cite{supp}).
In TSSE experiments, the typical width of the Pt contact $w=0.1~\hbox{mm}$,
and the length of the YIG $L=5~\hbox{mm}$ (see Tables~\ref{table1} and~\ref{table2}). In our simulations, we choose
a fixed ratio $L/w=32$ close to the value in real measurements. A large-scale atomistic simulation of the YIG$|$Pt system for real TSSE experiments ($w\sim 10^5 a$) is computationally too expensive.
Instead, we perform the calculation by systematically increasing
$w$ from $2a$ to $16a$ but fixing both the ratio of $\xi/w$ and the
temperature difference $\Delta T$. Through this approach, we
are able to quantitatively compare our simulations with
the experimental results.

\begin{figure}[ht!]
\centering
  \includegraphics[width=0.44\textwidth]{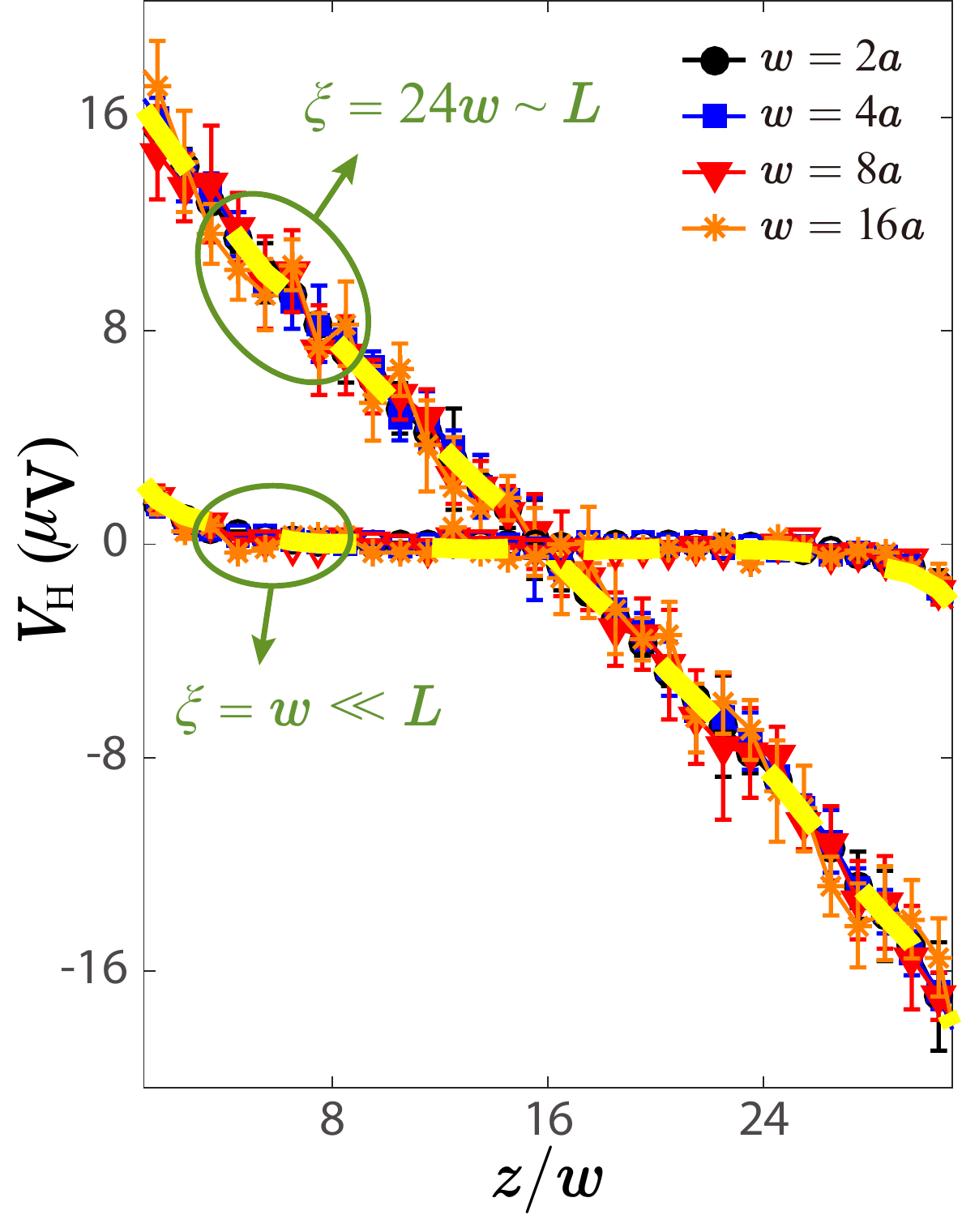}
  \caption{Inverse spin Hall voltage $V_H(z)$ for
  $w=2a,4a,8a,16a$ in 1D model. There are two bundles of curves,
  corresponding to the spread spin waves ($\xi=24w\sim L$) and the
  highly-localized spin waves ($\xi=w\ll L$). The localization length is
  tuned by the disorder strength through $\xi=(96J^2/W^2)a$. Yellow dash curves are the theoretical formula Eq.~(\ref{current}).
  }
\label{fig_sse}
\end{figure}

We first calculate the quasi 1D case. Figure~\ref{fig_sse}
displays the spatial distribution of
$V_H(z)$ for $\xi=24w$ $(\sim L)$ and $\xi=w$ $(\ll L)$ with different
Pt contact widths $w=2a,4a,8a,16a$, where symbols are numerical results.
The localization lengths $\xi$ are tuned by the disorder strength $W$ via the formula $\xi=96J^2a/W^2$ \cite{kramer1}. It shows that the voltage solely depends on the
ratio $\xi/w$, rather than the absolute value of the width
of the Pt contact. Note that it is difficult to estimate
or measure $\xi$ in real experiments, we
consider the case $\xi=24w=2.4~\hbox{mm}$ (corresponding to $W/J=7\times 10^{-3}$, or $W=0.14~\hbox{K}$ in real unit) in the analytical calculation, and find that
$V_H$ varies almost linearly with $z$ (see the upper yellow dashed curve in Fig.~\ref{fig_sse}). The
slope of $V_H(z)$ obtained both numerically and analytically is $6.4~\mu\hbox{V/mm}$, one order of magnitude larger than the experimental result ($0.5~\mu\hbox{V/mm}$) \cite{Uchida1}. Clearly, for the spread spin waves with
localization lengths approaching the system size, we
observe the sign change of the inverse spin Hall voltage at
high-/low-temperature regimes, namely the TSSE. However, a shut down of the TSSE sets in when the spin
wave localization length is of the same order of the width of the Pt contact, e.g., $\xi=w$, consistent with the theoretical analysis above. Numerical data compare well with the theoretical prediction (\ref{current})
[see the lower bundle of curves in Fig.~\ref{fig_sse}].
\par

\begin{figure}[ht!]
\centering
  \includegraphics[width=0.44\textwidth]{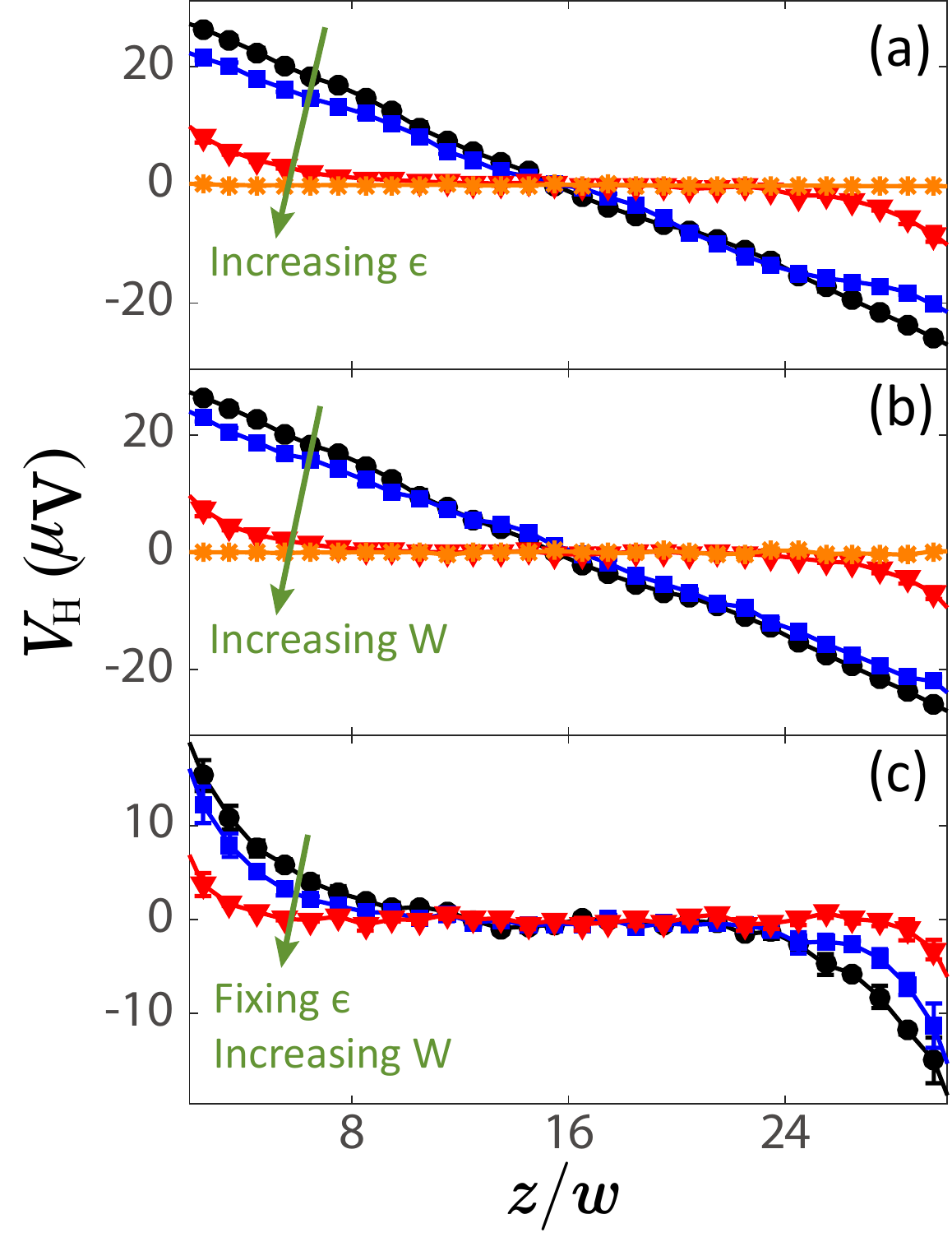}
  \caption{$V_H(z)$ for (a) $\epsilon=0,0.01,0.1,1$, and $W=0$ (Wannier-Zeeman
  localizations); (b) $\epsilon=0$, and $W=0,1,3,10$ (Anderson localizations);
  (c) $\epsilon=0.01$, and $W=0,1,3$ (both Wannier-Zeeman and Anderson localizations).
  The width of Pt contact is $w=4a$.
  }
\label{fig_sse1}
\end{figure}

For tunable Wannier-Zeeman localization (zero disorder, finite field gradient) and Anderson
localization (finite disorder, zero field gradient), the spatial dependence of the inverse spin Hall
voltage $V_H(z)$ is shown in Figs.~\ref{fig_sse1}(a) and \ref{fig_sse1}(b), respectively. In the calculations, we choose $w=4a$, without loss of generality, since it has been shown that TSSE only
depends on the ratio $\xi/w$.
In both cases, the TSSE is shown to be significantly suppressed by decreasing the localization length, and almost vanishes
if $\xi<w$ [see the orange lines in Figs.~\ref{fig_sse1}(a) and \ref{fig_sse1}(b)]. In
real YIG$|$Pt systems under the temperature gradient, disorders and
field gradients coexist, which should further suppress the TSSE. It
is indeed the case as shown in Fig.~\ref{fig_sse1}(c).
\par

\begin{figure}[ht!]
\centering
  \includegraphics[width=0.48\textwidth]{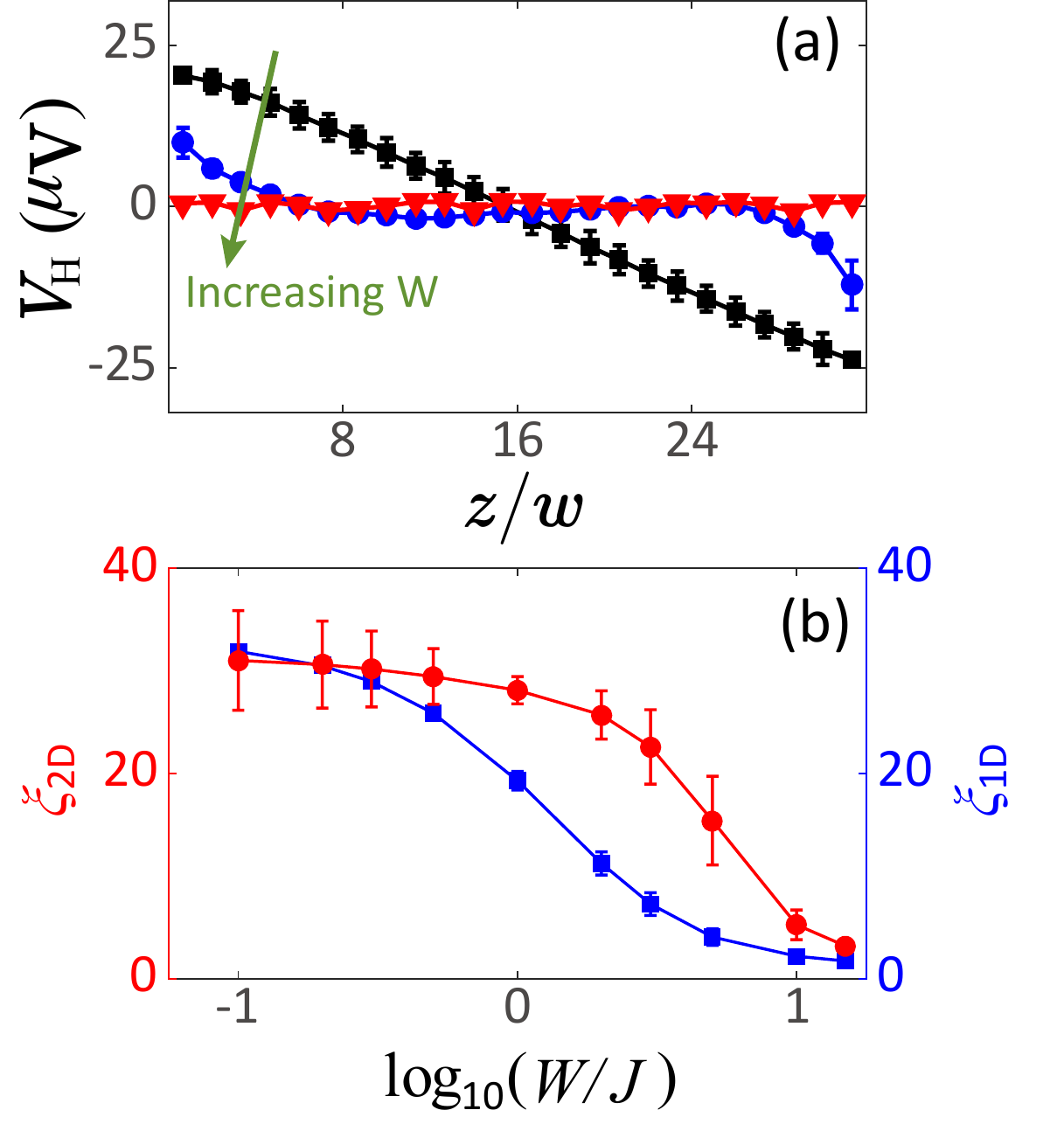}
  \caption{(a) Inverse spin Hall voltage $V_H(z)$ for $W=1$ (black squares),
  10 (blue circles), and 15 (red triangles) in 2D model. The length and width of the
  2D lattice are $L=64a$ and $L'=24a$ (the ratio $L/L'$ is close to the experiment value \cite{Uchida1}). The width of the Pt contact is $w=2a$.
  (b) Average participation number (square root of participation number)
  $\xi_{1D}$ ($\xi_{2D}$) as a function of $W$ for 1D (2D) system.
  The system sizes are chosen as $L^d$ for $L=48a$ and $d=1,2$.
  }
\label{fig_2d}
\end{figure}

Next, we proceed with the 2D case. In Fig.~\ref{fig_2d}(a), we plot
the $V_H$ versus the position $z$. It again shows that the
magnitude of inverse spin Hall voltage is strongly suppressed
by increasing the disorder, similar to what we observed in
1D system above. By comparing the results of 1D and 2D, cf. Fig. \ref{fig_sse1}(b) and Fig.~\ref{fig_2d}(a), we find that the magnitude of the TSSE in 2D is relatively larger than that in 1D (but still in the same order). This can be explained by the fact that the localization length in high dimension is usually larger than that in the low dimension, which indeed can be seen in Fig.~\ref{fig_2d}(b) by evaluating the
average participation numbers of both 1D and 2D systems [see Methods~\ref{method-2}
for more details]. Moreover, we expect that the spin waves can be
cooperatively localized in 2D system with even weaker disorders
assisted by the field gradient (See Supplemental Material Sec.~II
\cite{supp}), just as the 1D results shown in Fig. \ref{fig_sse1}(c). Our study thus quantitatively validates the role of spin-wave Anderson (Wannier-Zeeman)
localization in TSSE, regardless of the system dimension.

\section{Wave turbulence}

\begin{figure}[ht!]
\centering
  \includegraphics[width=0.44\textwidth]{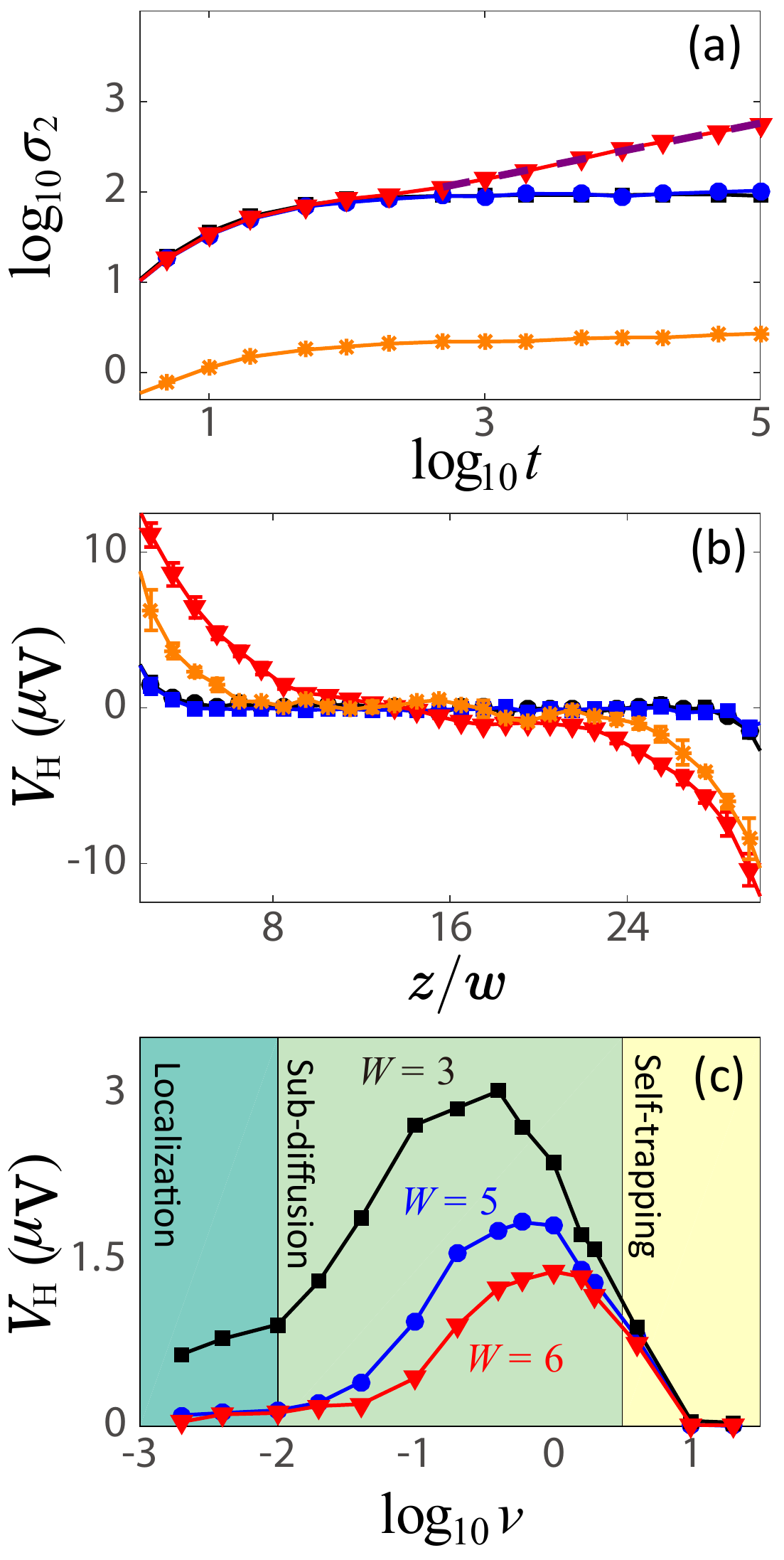}
  \caption{(a) Wave-packet second-moment $\log_{10}\sigma_2$ v.s. $\log_{10}t$
  for $\nu=0$ (black squares), 0.01 (blue circles), 1 (red triangles), and
  10 (orange stars) at $\epsilon=0$ and $W=3$. Values of $\log_{10}\sigma_2$
  are averaged over time interval $\Delta(\log_{10}t)\simeq 0.1$. Purple dashed
  line is the linear fit of $\log_{10}\sigma_2=\beta\log_{10} t+\gamma$ with
  $\beta=0.306\pm0.002$ and $\gamma=1.22\pm0.01$.
  (b) Spatial distribution of $V_H$ for $\epsilon=0,W=3$, $\nu=0$ (black squares),
  0.001 (blue circles), 0.1 (red triangles), and 1 (orange stars).
  (c) $V_H(z=4w)$ v.s. $\log_{10}\nu$
  for $\epsilon=0$ under different disorder strengths. $w=4a$ is used in (b) and (c).
  }
\label{fig_turbulence}
\end{figure}

Remarkably, the spin-wave turbulence due to many-body interactions enriches
the above picture because a large enough nonlinearity tends to delocalize
magnons after a long run (reaching a steady state). There are three relevant energy scales: the average
energy spacing $\overline{\delta \omega}$, the spectrum band-width $\Delta$,
and the energy shift due to the magnon-magnon interaction $g \sim \nu|\psi_{n}|^2$.
One expects three different regimes: (i) $g<\overline{\delta\omega}$
where the nonlinearity is not strong enough to cause the wave turbulence, and
spin waves are localized; (ii) $\overline{\delta\omega}<g< \Delta$ where
all spin waves are in turbulent states and delocalized;
(iii) $\Delta<g$ where the nonlinearity is so strong that spin waves
fall into the self-trapping region. To confirm this criterion,
we study the spin wave spreading at $T=0~\hbox{K}$ in 1D system
by numerically calculating the time evolution of the second-moment
$\sigma_2(t)\equiv\langle\sum_n (n-\langle n\rangle)^2|\psi_n(t)|^2\rangle$
of an initial single-site excitation for different nonlinearities, shown in
Fig.~\ref{fig_turbulence}(a).
Indeed, the spin waves undergo a {\it sub-diffusive} motion
$\sigma_2\propto t^{\beta }$ for a strong nonlinearity $\nu=1$ while the
Anderson localization occurs for $\nu=0$ and $0.001$.
Our numerical data
indicates $\beta =0.306\pm0.002$, which is consistent with Ref.~\cite{Pikovsky}.
For a very large nonlinearity $\nu=10$, spin waves are self-trapped.
\par

The TSSE then is expected to follow the predictions in the three different regions. Figure~\ref{fig_turbulence}(b) shows the spatial dependence of inverse
spin Hall voltage for a fixed disorder $W=3$ under different nonlinearities $\nu=0,0.001,0.1$
and $1$ in 1D. Clearly, strong nonlinearities ($\nu=0.1$ and 1) can significantly enhance the
TSSE due to the delocalization of magnons, compared to the results of $\nu=0$ and $0.001$. Typically, the nonlinearity is about $\nu=0.001$ for pure YIG (see Table~\ref{table2} in Appendix~\ref{app3}). Thus, the
YIG$|$Pt systems in most TSSE experiments belong to the Anderson localization
phase, where the TSSE signals are too small to be observed. However, for the
doped YIG (YBi)$_3$(FeGa)$_5$O$_{12}$ with high anisotropies \cite{novoselov},
$\nu=0.84$ (or 16.8~\hbox{K} in real unit), one may expect a transition
from the Anderson localization to the sub-diffusion, thus a giant
enhancement of the TSSE. However, the TSSE is suppressed again
by an ultrastrong nonlinearity due to the magnon self-trapping.
Figure~\ref{fig_turbulence}(c) shows the predicted three phases of
TSSE.
\par

Finally, we would like to point out that, at elevated temperature, the spin wave dispersion relation is shifted due to
the nonlinearity. For extended spin waves, we are able to analytically calculate the
dispersion of the renormalized mode due to the magnon-magnon
interaction (See Appendix~\ref{app2} for the derivation),
\begin{equation}
    \begin{gathered}
        \tilde{\omega}_k=\omega_k+\dfrac{2\nu\gamma}{S}\sum^N_{p=1} \bar{n}_p /N,
    \end{gathered}\label{wide1}
\end{equation}
where $\bar{n}_{p}$ is the number of the $p$th normal magnon mode
at steady states. Numerical results excellently agree with the analytical formula \eqref{wide1} (See Supplemental Material Sec.~III \cite{supp}).
For YIG at room temperature ($300~\hbox{K}$), we estimate
the energy shift due to the renormalization as $0.536~\hbox{K}$,
which can not be neglected since it is comparable
to the temperature difference applied at the two ends of YIG in the
TSSE experiment \cite{Uchida1}.
Naturally, the renormalized spin wave should act as the carrier for both the heat and
the spin currents in TSSE, which is numerically demonstrated in Supplemental Material Sec.~III \cite{supp}.
\par

\section{Discussion}

We would like to make a few more remarks before concluding this 
article. (i) In principle, our theory
should hold for the SSE in both transverse and longitudinal configurations.
(ii) The electric signal observed in TSSE experiments comes not only from
SSE but inevitably also from other sources such as the anomalous Nernst
effect \cite{Chien1} and the conventional Seebeck effect driven by
thermal Hall current in ferromagnets, as well as the longitudinal SSE
due to the local temperature differences between YIG and Pt bar.
Although the TSSE signal is sensitive to spin-wave localization,
other sources do not require delocalization of spin waves. Thus,
one can experimentally separate a TSSE signal from other signals by
applying (or not applying) a large field gradient along $z-$direction
to switch off (on) TSSE signal through the Wannier-Zeeman localization.
For the YIG$|$Pt system with $w=0.1~\hbox{mm}$, a magnetic field gradient
$\epsilon=0.21~\hbox{T/mm}$ is able to switch off the TSSE response (see Methods~\ref{method-3}).
Taking into account the effect of both spin-wave Anderson
and Wannier-Zeeman localizations, a much smaller field-gradient should
be enough for that purpose. (iii) Interestingly, we note a long-range spin transport in disordered
magnets recently reported in the amorphous YIG \cite{Wesenberg1}. The
model studied here is microscopically relevant to such experiments. (iv) How to generalize the present study to 3D and
antiferromagnets should be interesting subjects, and also poses both theoretical
and numerical challenges.
\par

\section{Conclusion}

In this work, we reveal the relationship between the spin-wave localization and the
TSSE.
By carefully tuning the localization length $\xi$ through either the Anderson or the Wannier-Zeeman mechanism, we are able to identify that
spread spin waves ($\xi\sim L$) in a ferromagnetic insulator (YIG) and
electrons in a non-magnetic heavy metal (Pt) generally are not at thermal equilibrium (except at one single point where there is no spin pumping), while highly localized spin waves ($\xi\sim w$) and electrons are always at
thermal equilibrium.
This naturally explains why spin current pumped from ferromagnetic insulator
to the detecting bar reverses its direction at the high-/low-temperature side, and gives a route to understand why different experiments can show contradictory results.
Furthermore, we show that the nonlinearity tends
to delocalize spin waves in the weakly turbulent region and thus enhances
the TSSE, but leads to spin wave self-trapping in the strongly turbulent
region so that the TSSE is significantly suppressed again. Our theory provides a novel framework to understand the tantalizing spin Seebeck effect and to resolve the lingering debate on the very existence of the TSSE, with clear experiment scheme proposed to test it.
\par

\section{Methods}

\subsection{Atomistic simulation of spin dynamics}
\label{method-1}

We adopt the modified SBAB$_2$ method to numerically calculate the spin wave
functions $\psi_{\bm{n}}(t)$ governed by Eq.~\eqref{DNLS}. The SBAB$_2$
method is an effective scheme to simulate the time evolution of the
discrete nonlinear Schr\"{o}dinger equation. Conventionally, the
dissipations and thermal fluctuations are not included in the SBAB$_2$
method (see Ref.~\cite{Pikovsky} and references therein).
We generalize this method to include the thermal effect by using the It\^{o}
stochastic calculus, and numerically calculate the inverse spin Hall
voltage by Eq.~\eqref{current1} (see Supplemental Material Sec.~I \cite{supp}).
The accuracy of the modified SBAB$_2$ method has been tested:
We numerically reproduce the well-known energy
equipartition for equilibrium systems ($\Delta T=0$), as well as the
energy repartition \cite{Yan1} for systems at nonequilibrium steady
states in Supplemental Material Sec.~III \cite{supp}.
\par

\subsection{Participation numbers}
\label{method-2}

To quantify the degree of localization of spin waves in finite-size
disordered YIG, we numerically compute the participation number,
defined as \cite{Wang123}
\begin{equation}
\begin{gathered}
PN(\omega)=\dfrac{1}{\sum_{\bm{n}}|\psi_{\bm{n}}(\omega)|^4},
\end{gathered}\label{pn}
\end{equation}
by exactly calculating the $\psi_{\bm{n}}(\omega)$ corresponding to an eigenfrequency
$\omega$ at $T=0~\hbox{K}$. For a maximal spread (uniform) state, we have $\psi_{\bm{n}}=1/\sqrt{N}$ $\forall \bm{n}$ and $PN=N$; while for a
maximal localized state (single site excitation), we have $PN=1$. Thus, a larger
$PN$ indicates a less localized spin wave. We use the
average participation number, $\xi_{1D}=(1/N)\sum_{\omega}PN(\omega)$ to
measure the extendedness of spin waves in 1D systems; while for 2D systems,
we use the square root of the average participation number,
$\xi_{2D}=\sqrt{(1/N)\sum_{\omega}PN(\omega)}$.
\par

\subsection{Estimation of the magnetic field gradient to switch off TSSE}
\label{method-3}

For Wannier-Zeeman localization, the magnon eigenfunction is the
Bessel function of the first kind with asymptotic localization length
$\xi\simeq -a/[(\epsilon/J)\ln(\epsilon/J)]$ for $\epsilon\rightarrow 0$
in a quasi-1D spin chain \cite{Yan1}. When $\xi<w$, the TSSE signal is
switched off. Typically, the width of the detecting metal contact is
$0.1~\hbox{mm}$. To localize spin waves in
such length scale through the Wannier-Zeeman mechanism, one needs the
on-site energy $K=1.44~\hbox{K}$ at one end and $0$ at the other end,
for instance. For a system of length $L=5~\hbox{mm}$, the required magnetic
field gradient is $0.21~\hbox{T/mm}$.
\par

\begin{acknowledgments}

This work is supported by the National Natural Science Foundation of China
under Grant No.~11604041 and  No. 11704060, the National Thousand-Young-Talent
Program of China. X.R.W. is supported by National Natural Science Foundation
of China (Grant No.~11374249) and Hong Kong RGC (Grants No.~16301115 and
16301816). C.W. is supported by the China Postdoctoral Science Foundation
(Grants No.~2017M610595 and 2017T100684) and the National Nature Science
Foundation of China under Grant No.~11704061.
\par

\end{acknowledgments}
%\newpage
\appendix

\section{Analytical expression for magnon-induced spin injection in disordered YIG$|$Pt systems}
\label{app1}

By modelling the disordered YIG as a 1D spin chain, we are able to analytically
derive the net spin current pumped from YIG into Pt contact. In the absence of the dissipation,
thermal fluctuation, and nonlinearity, Eq.~\eqref{DNLS} reduces to
\begin{equation}
\begin{gathered}
i\dot{\psi}_n=\mathcal{H}_0\psi_n=\dfrac{\gamma}{S}[J(\psi_{n+1}+\psi_{n-1})-K_n\psi_n]\\
=-\omega_k\psi_n,
\end{gathered}\label{app_1_1}
\end{equation}
where $\omega_k$ is the frequency of the linear Hamiltonian $\mathcal{H}_0$, and
$k$ labels the mode index. $\mathcal{H}_0$ can be diagonalized by an
orthogonal matrix $P$, i.e., an expansion of $\psi_n$ into
its mode space $\psi_n=\sum_k P_{nk}\phi_k$. When taking the thermal noise $\Theta_n$ into account, we have
\begin{equation}
\begin{gathered}
(i+\alpha)\dot{\phi}_k=-\omega_k\phi_k+\Lambda_k,
\end{gathered}\label{app_1_2}
\end{equation}
where $\Lambda_k=\sum_n P_{nk}\Theta_n$. The DC component of spin current pumped
from YIG to Pt can be calculated by Eq.~\eqref{current1}
\begin{equation}
\begin{gathered}
j^z_{s,n}
=-\dfrac{g^{\uparrow\downarrow}_{\text{eff}}\hbar}{4\pi}
\text{Im}\left[  \sum_{kk'} P_{nk}P_{nk'} \Bigg\langle\phi_k\dot{\phi}_{k'}
\Bigg\rangle\right],
\end{gathered}\label{app_1_3}
\end{equation}
where $\langle\cdots\rangle$ represents the time averaging.
$\langle\phi_k\dot{\phi}_{k'}\rangle$ can be obtained by Eq.~\eqref{app_1_2}
\begin{equation}
\begin{gathered}
\langle \phi_k \dot{\phi}^{\ast}_{k'}\rangle = \dfrac{i+\alpha}{1+\alpha^2}
\left(-\omega_{k'}\langle\phi_{k}\phi^{\ast}_{k'}\rangle
+\langle \Lambda^{\ast}_{k'}\phi_k \rangle\right).
\end{gathered}\label{app_1_4}
\end{equation}
For most TSSE experiments, the time scale of the dynamics of magnetizations
$\tau_m$ is much longer than the time scale of the noisy environment
$\tau_{n}$ such that $\Lambda^{\ast}_{k'}$ and $\phi_k$ can be treated
as independent variables. Thus, the second term in the right-hand side
of Eq.~\eqref{app_1_4} vanishes,
\begin{equation}
\begin{gathered}
\langle \Lambda^{\ast}_{k'}\phi_k \rangle = \langle \Lambda^{\ast}_{k'}\rangle
\langle\phi_k \rangle=0.
\end{gathered}\label{app_1_5}
\end{equation}
To evaluate the first term of Eq.~\eqref{app_1_4}, we integrate
Eq.~\eqref{app_1_2} from $t$ to $t+\tau$ with $\tau_{n}\ll\tau\ll\tau_m$,
\begin{equation}
\begin{gathered}
\phi_k(t+\tau)=\phi_k(t)+
(i-\alpha)\omega_k\phi_k(t)\tau \\
-(i-\alpha)\int^{t+\tau}_t\Lambda_k(t_1)dt_1.
\end{gathered}\label{app_1_6}
\end{equation}
Thus, we have
\begin{equation}
\begin{gathered}
\dfrac{d\langle\phi_k\phi^{\ast}_{k'}\rangle}{dt}=-\dfrac{1}{1+\alpha^2}
\left[(\alpha(\omega_k+\omega_{k'})\right.\\
\left. +i(\omega_k-\omega_{k'}))
\langle\phi_k\phi^{\ast}_{k'}\rangle
+4\alpha k_B \gamma \mathcal{T}_{kk'}/(S)\right],
\end{gathered}\label{app_1_7}
\end{equation}
where $\mathcal{T}_{kk'}=\sum_n P_{nk}P_{nk'}T_n$. At steady state
$d\langle\phi_k\phi^{\ast}_{k'}\rangle/dt=0$,
\begin{equation}
\begin{gathered}
\langle\phi_k\phi^{\ast}_{k'}\rangle=\dfrac{4\alpha k_B \mathcal{T}_{kk'}\gamma}
{[\alpha(\omega_k+\omega_{k'})+i(\omega_k-\omega_{k'})]S}.
\end{gathered}\label{app_1_8}
\end{equation}
Substituting Eq.~\eqref{app_1_8} into Eq.~\eqref{app_1_3}, we obtain the
expression of the net spin current,
\begin{equation}
\begin{gathered}
j^z_{s,n}=\dfrac{2\alpha^2 g^{\uparrow\downarrow}_{\text{eff}}\hbar\gamma}
{\pi(1+\alpha^2)S}\sum_{kk'}P_{nk}P_{nk'}\\
\times\dfrac{k_B(\mathcal{T}_{kk'}-T_n)\omega_k
\omega_{k'}}{\alpha^2(\omega_k+\omega_{k'})^2+(\omega_k-\omega_{k'})^2},
\end{gathered}\label{app_1_9}
\end{equation}
where we include the Johnson-Nyquist noise generated in the Pt contact that is
proportional to the electron temperature $T_e$, equal to the local temperature of the thermal bath. Disregarding the Kapitza interface heat
resistance, the temperature of YIG is continuous over the interface, and
$T_e=T_n$. For YIG with a small Gilbert damping $\alpha=10^{-4}$, only the
diagonal terms in Eq.~\eqref{app_1_9} contribute to the net spin current,
\begin{equation}
\begin{gathered}
j^z_{s,n}=\dfrac{g^{\uparrow\downarrow}_{\text{eff}}\hbar\gamma}
{2\pi S}k_B\left(\sum_k (P_{nk})^2\mathcal{T}_k-T_n\right),
\end{gathered}\label{app_1_10}
\end{equation}
where $\mathcal{T}_k=\sum_n (P_{nk})^2 T_n$ can be regarded as the
temperature of the $k$th mode. Eq.~\eqref{app_1_10} is
Eq.~\eqref{current} in the main text.
\par

The net spin current $j^z_{s,n}$ pumped from the YIG into the Pt
contact can be detected by the inverse spin Hall effect. The net
spin current at site $n$ gives rise to a DC Hall current $j^y_n$,
\begin{equation}
\begin{gathered}
j^y_n=\theta_H\dfrac{2e}{\hbar}j^z_{s,n},
\end{gathered}\label{app_1_11}
\end{equation}
where $\theta_H$ is the inverse Hall angle. The length and width
of the Pt contact are $L'$ and $w$, respectively. The electric
voltage over the two transverse ends of the Pt contact with position
$z$ can be calculated by
\begin{equation}
\begin{gathered}
V_H(z/w)=\dfrac{2\rho e L'\theta_H}{A\hbar}\overline{j}^z_{s}(z/w).
\end{gathered}\label{app_1_12}
\end{equation}
Here $\rho$ is the resistivity of the Pt contact, and $A$ is the
contact area. $\overline{j}^z_{s}(z/w)$ is the mean net spin current pumped
into the Pt contact.

\section{Renormalized dispersion relation}
\label{app2}

In the mode space of 1D model, Eq.~\eqref{DNLS} reads
\begin{equation}
\begin{gathered}
(i+\alpha)\dot{\phi}_k=-\omega_k\phi_k\\
-\dfrac{\nu\gamma}{S}\sum_{k_1 k_2 k_3} I_{k k_1 k_2 k_3}
\phi_{k_1}\phi^{\ast}_{k_2}\phi_{k_3}
+\Lambda_k,
\end{gathered}\label{app_2_0}
\end{equation}
where $I_{k k_1 k_2 k_3}=\sum_n P_{nk}P_{nk_1}P_{nk_2}P_{nk_3}$ represents
the coupling between different spin-wave modes due to the nonlinearity.
To investigate the nonlinear effect, we integrate Eq.~\eqref{app_2_0}
from $t$ to $t+\tau$ with $\tau_{n}\ll\tau\ll\tau_{m}$
\begin{equation}
\begin{gathered}
\phi_k(t+\tau)=\phi_k(t)+(i-\alpha)\omega_k\phi_k(t)\tau\\
+(i-\alpha)\dfrac{\nu\gamma}{S}\sum_{k_1 k_2 k_3}I_{k k_1 k_2 k_3}\phi_{k_1}\phi^{\ast}_{k_2}
\phi_{k_3}\\
-(i-\alpha)\int^{t+\tau}_t\Lambda_k(t_1)dt_1.
\end{gathered}\label{app_2_1}
\end{equation}
Then, we are able to find the time evolution equation of the mean
number of the $k$th mode, $n_k=\langle\phi_k\phi^{\ast}_k\rangle$,
\begin{equation}
\begin{gathered}
    \dot{n}_k=-2\alpha\omega_k n_k+4\alpha k_B \mathcal{T}_k\gamma/S\\
    -2(\nu\gamma/S)\sum_{k_1 k_2 k_3}
    I_{kk_1k_2k_3}\text{Re}
    \left[(\alpha-i)\langle
    \phi^{\ast}_k\phi_{k_1}\phi^{\ast}_{k_2}
    \phi_{k_3}\rangle\right].
\end{gathered}\label{app_2_2}
\end{equation}
We apply the fixed boundary conditions at two ends such that the number
of spins reduces to $N-1$ effectively. The eigenvalues of the linear
Hamiltonian $\mathcal{H}_0$ are
\begin{equation}
\begin{gathered}
    \omega_k=2(J\gamma/S)[1-\cos(k\pi/N)],
\end{gathered}\label{app_2_3}
\end{equation}
and the eigenfunctions are
\begin{equation}
\begin{gathered}
    P_{nk}=\sqrt{2/N}\sin(nk\pi/N),n=1,2,\cdots,N-1.
\end{gathered}\label{app_2_4}
\end{equation}
Under the resonant condition, say $k=k_1$ and $k_2=k_3$, the coupling
integral reads
\begin{equation}
\begin{gathered}
    I_{k k_1 k_2 k_3}=\left\{
    \begin{array}{c}
    1/N\text{, for }k\neq k_2\text{ or }N-k_2 \\
    \quad\\
    3/(2N)\text{, for }k = k_2\text{ or }N-k_2 \\
    \end{array}
    \right..
\end{gathered}\label{app_2_5}
\end{equation}

\begin{widetext}

For a moderate nonlinearity, the number of magnons approximately follows the
Gaussian distribution. By using Wick’s theorem, we have
\begin{equation}
\begin{gathered}
    \langle\phi_k\phi^{\ast}_k\phi_k\phi^{\ast}_k\rangle=
    \langle\phi_k\phi^{\ast}_k\rangle\langle\phi_k\phi^{\ast}_k\rangle
    + \langle\phi^{\ast}_k\phi_k\rangle\langle\phi^{\ast}_k\phi_k\rangle.
\end{gathered}\label{app_2_6}
\end{equation}
Thus, the nonlinear term in Eq.~\eqref{app_2_2} reads
\begin{equation}
\begin{gathered}
    \text{Re}\left[\dfrac{4\nu\tau}{i+\alpha}\sum_p I_{kkpp}
    \langle\phi^{\ast}_k\phi_k\phi_p\phi_p\rangle\right]\\
    =
    \dfrac{4\nu\tau\alpha}{N}\left(\sum_p n_p+2n_k+\dfrac{1}{2}n_{N-k}\right)n_k.
\end{gathered}\label{app_2_7}
\end{equation}
In the thermodynamic limit, i.e., $N\rightarrow\infty$,
\begin{equation}
    \begin{gathered}
        \dot{n}_k=-2\alpha \omega_k n_k+\dfrac{\gamma}{S}\left[4\alpha k_B \mathcal{T}_k-4\nu\alpha
        \left( \sum^N_{p=1} n_p /N \right) n_k\right].
    \end{gathered}\label{app_2_8}
\end{equation}
Equation~\eqref{app_2_8} demonstrates three important features:
(i) For extended spin waves, every mode interacts with all the other modes.
(ii) The energy repartition for nonequilibrium steady states, i.e.,
$E_k=\bar{n}_k\tilde{\omega}_kS/(2\gamma)=k_B \mathcal{T}_k$ with
$\bar{n}_k\equiv n_k(t\rightarrow\infty)$, is exactly satisfied by
the {\it renormalized mode} Eq.~\eqref{wide1}.
Thus, the nonlinearity makes a mode-independent frequency shift
$\Delta\omega=\tilde{\omega}_k-\omega_k$. The renormalized dispersion
shifts upward (downward) for a positive (negative) $\nu$;
(iii) The steady distribution $\bar{n}_k$ may go to infinite for
a strong enough attractive magnon-magnon interaction ($\nu<0$),
and leads to an instability.

\section{Parameters for the YIG$|$Pt system}
\label{app3}

\begin{table*}[!ht]
\caption{\label{table2} Materials parameters.}
\begin{ruledtabular}
\begin{tabular}{ccc}
\multicolumn{3}{c}{YIG} \\
Quantity & Values & References \\
\hline
Gyromagnetic ratio $\gamma$ & $1.76\times10^{11}~\hbox{rad}/\hbox{Ts}$ & \\
Gilbert damping $\alpha$ & $10^{-4}$ & \cite{Kajiwara1} \\
Saturation magnetization $ 4\pi S/a^{3}$ & $1.4\times 10^5~\hbox{A/m}$ & \cite{Kajiwara1} \\
Exchange energy $J$ & $20~\hbox{K}$ & \cite{Shinozaki1,Xie1} \\
Nonlinearity $\nu$ & $0.02~\hbox{K}$ &  \cite{Hannes1} \\
Mixing conductance $g^{\uparrow\downarrow}_{\text{eff}}/A$ ($A$ the contact area) & $1\times10^{11} \hbox{cm}^{-2}$
& \cite{Uchida1} \\
External magnetic field $B$  & $0.32~\hbox{T}$ & \cite{Uchida1} \\
& & \\
\multicolumn{3}{c}{Pt contact} \\
Quantity & Values & References \\
\hline
Hall angle $\theta_H$ & 0.00037 & \cite{Kimura1} \\
Resistance $\rho$ & $1\times 10^{-6}~\Omega\hbox{m}$ & \cite{Uchida1} \\
Geometry $L'\times w\times h$ & $4~\hbox{mm}\times 0.1~\hbox{mm}\times 15~\hbox{nm}$
& \cite{Uchida1} \\
\end{tabular}
\end{ruledtabular}
\end{table*}

\end{widetext}

%\bibliography{References}

\end{document}